%% file: paper.tex
\documentclass[journal]{vgtc}                
\usepackage{mathptmx}
\usepackage{color}
\usepackage{graphicx}
\usepackage{algorithm}
\usepackage{algpseudocode}
\usepackage{times}
\usepackage{booktabs}
\usepackage[bookmarks,backref=true,linkcolor=black]{hyperref} 
\hypersetup{
  pdfauthor = {},
  pdftitle = {},
  pdfsubject = {},
  pdfkeywords = {},
  colorlinks=true,
  linkcolor= black,
  citecolor= black,
  pageanchor=true,
  urlcolor = black,
  plainpages = false,
  linktocpage
}

\onlineid{0}

\vgtccategory{Research}

\vgtcinsertpkg

\title{Visual Model Validation via Inline Replication}

\author{David Gotz, Brandon A. Price, and Annie T. Chen}
\authorfooter{
\item David Gotz is with the University of North Carolina at Chapel Hill. E-mail: gotz@unc.edu.
\item Brandon A. Price is with the University of North Carolina at Chapel Hill. E-mail: baprice@live.unc.edu.
\item Annie T. Chen is with the University of Washington. E-mail: atchen@uw.edu.
}

\shortauthortitle{Gotz, Price, and Chen: Visual Analysis and Interactive Replication}
\abstract{
Data visualizations typically show retrospective views of an existing dataset
with little or no focus on repeatability.  However, consumers of these tools often use
insights gleaned from retrospective visualizations as the basis for decisions
about future events.  In this way, visualizations often serve as visual
predictive models despite the fact that they are typically designed to present
historical views of the data.  This ``visual predictive model'' approach, however, can lead to invalid inferences.  In this
paper, we describe an approach to visual model validation called \emph{Inline
Replication} (IR) which, similar to the cross-validation technique used widely
in machine learning, provides a nonparametric and broadly applicable technique for
visual model assessment and repeatability.  This paper describes the overall IR process and
outlines how it can be integrated into both traditional and emerging ``big
data'' visualization pipelines.  Examples are provided showing IR integrated
within common visualization techniques (such as bar charts and linear
regression lines) as well as a more fully-featured visualization system
designed for complex exploratory analysis tasks.
}


  \teaser{
  \centering
  \includegraphics[width=12.8cm]{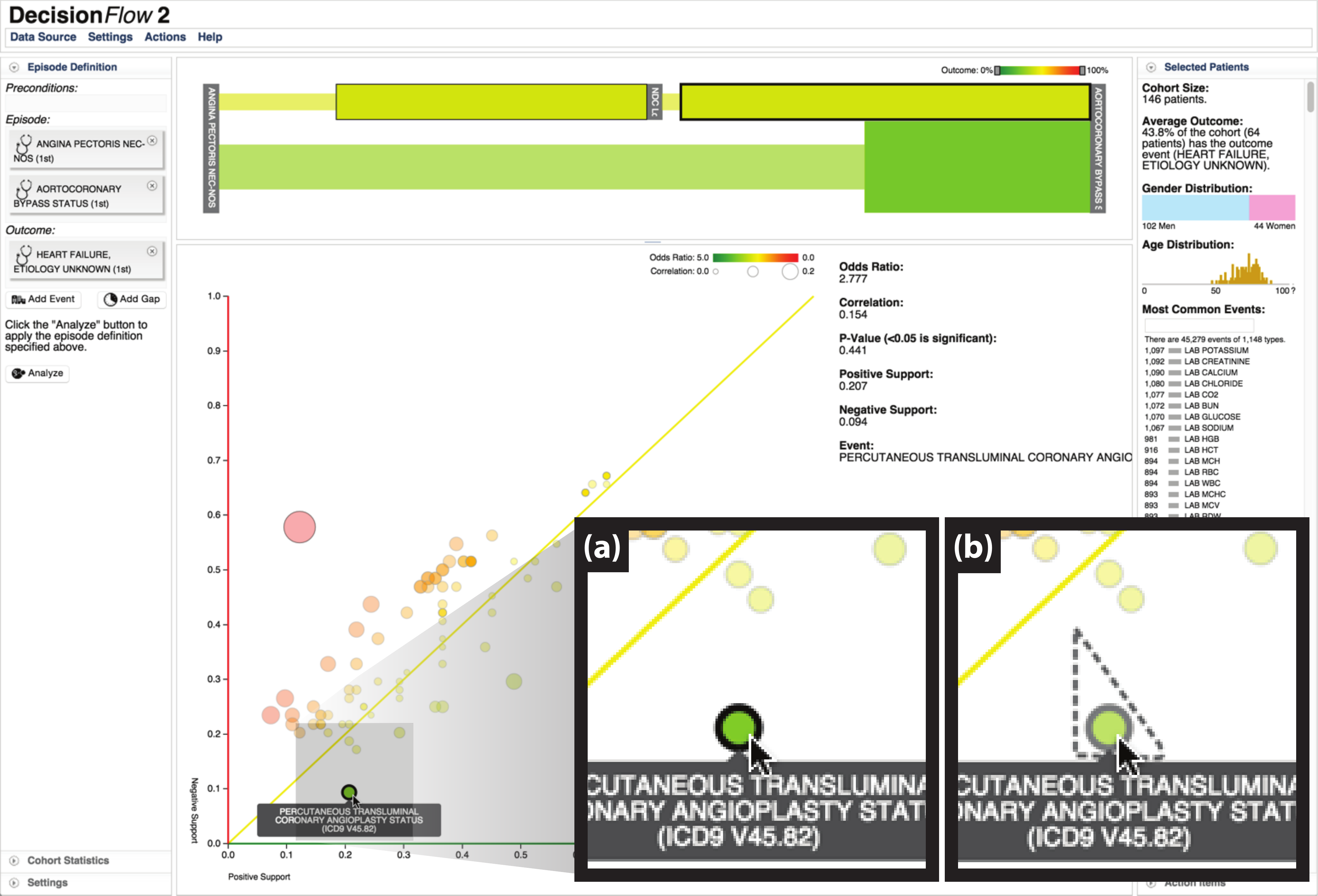}
  \vspace{0.1cm}
  \caption{The DecisionFlow2 visual analytics system is shown here displaying medical event data using the Inline Replication (IR) process outlined in this paper.  The data shown in this example has been analyzed using five folds, without replacement.  The insert subfigures show (a) an initial visualization of the aggregation function's results for a particular medical event, and (b) a more detailed ``unfolded'' representation showing the variation in positive and negative support as observed across the five folds produced by the partition function.}
  \label{fig:teaser}
  \vspace{0.1cm}
}

\begin{document}

\maketitle

\input{intro}

\input{background}

\input{prediction}

\input{theory}
\input{usecase}

\input{conclusion}

\section{Acknowledgements}

This research was made possible, in part, by funds from a 2015 Data Fellow Award from the National Consortium for Data Science (NCDS).

\newpage
\newpage
\bibliographystyle{abbrv}
\bibliography{vast_pvalues.bib,vast_websites.bib}
\end{document}

%% file: intro.tex
\section{Introduction}
\label{sec:intro}

Visualizations are most often designed to depict the entirety of a
dataset--subject to a set of filters applied to focus the analysis--as
accurately as possible.  In this typical pattern, the goal is to provide a
person using the visualization with an accurate understanding of \emph{all of
the data} in the underlying dataset that matches the active set of filters.
This ethos was captured, perhaps most famously, in Shneiderman's Visual
Information Seeking Mantra: \emph{overview first, zoom and filter, then
details-on-demand} \cite{shneiderman_eyes_1996}.  Variations of this basic
approach have since been adopted in most modern visualization systems.  

The foundation for these systems are visual mappings that specify a graphical
representation for the underlying data.  For small and low-dimensional data
sources, these mappings can be direct (e.g., a scatter plot for a small two
dimensional dataset).  As problems grow in data size or dimensionality,
algorithmic data transformation methods can be used to filter, manipulate, and
summarize raw data into a more easily visualized form.

On top of these mappings, interactive controls are often provided to allow
users even more flexibility to filter or zoom to specific subsets of data.
These interactions can be linked to more detailed information about data
objects, for example via levels-of-detail or multiple coordinated views.  The
result, when well designed, is an effective visual interface for data
exploration and insight discovery.  

For this reason, these steps form the core stages of the canonical
visualization pipeline \cite{card_readings_1999, chi_taxonomy_2000}.
This approach can be enormously informative, and it has led
to revolutions in how people seek to understand information.  This approach
can be used, for example, to visualize file systems (showing the
space used by various directories) to help computer users navigate through large
directories; to visualize medical records to help doctors
understand patient histories; and to visualize maps of weather data
to identify regions most impacted by a recent storm.

Critically, however, these visualization use cases are all \emph{retrospective}
in nature.  Moreover, they describe visualizations that faithfully report data
as it was observed.  Users aim to see an overview of the entirety of a dataset.
If a user applies constraints to focus the visual investigation (e.g., via zoom
and filter), the visualization is expected to show the full set of data that
satisfies the applied constraints.

In many visualization scenarios, however, users are in fact more interested in
conducting \emph{prospective analysis}: using historical data to reason about
future or not-yet-observed data.  For example, medical experts examining data
for a cohort of patients might be most interested in what treatments would work
best for a future patient with similar characteristics.  Visualizations of
historical sports statistics are often used to inform strategic decisions that
are used in upcoming competitions.  Financial visualization tools are often
used to inform future investment decisions.  In each of these use cases,
visualizations of historical data are used to inform future decisions.  

For such prospective analysis tasks, retrospective visualizations are often
used as naive \emph{visual predictive models} with the assumption that
historical data can be predictive of future observations.  In fact, in many
cases retrospective representations are indeed very informative.  
However, just like the underlying descriptive statistics that such
visualizations often depict \cite{ostle_statistics_1963}, traditional
retrospective visualizations often provide insufficient evidence for making
predictive inferences.  

This critical gap between (a) retrospective visualization designs and (b) the
predictive requirements of many users has been recognized within the
visualization community \cite{perer_ieee_2014}.
Some have attempted to bridge this gap by adding support for 
inferential statistics within the visualization.  Typically, this approach
combines carefully designed statistical models with visualizations of the
model's results.  For example, visualizations can be instrumented to estimate
and display uncertainty, confidence intervals, or statistical significance.
Alternatively, predictive modeling methods can be used to generate additional
data, with the predictions themselves being incorporated into the
visualization.  These systems go beyond traditional descriptive reporting, but
they typically require a careful and sometimes onerous focus
on modeling, including estimates for underlying statistical distributions.

This paper presents \emph{Inline Replication} (IR), an alternative approach
to enabling inferential interpretation that is designed to overcome many of
the above challenges.  Our method, made practical by the ever-larger datasets now
available in many applications, is motivated by the cross-validation technique
used widely in machine learning.  The IR approach is nonparametric, making it
easy to apply and use generically within a visualization system without arduous 
modeling.  In addition, IR is ideal for use in large-scale visualization
systems where progressive or sample-based approaches are required. Finally, our
method provides users with validation information that is both intuitive and
easy to interpret.  

This remainder of this paper is organized as follows.  It begins with a review
of related work, then describes the details of the IR methodology.  We then
share example results from a variety of proof-of-concept systems that have
adopted the IR technique.  These examples range from simple bar charts to more
sophisticated interactive visualizations of large scale event data collections
\cite{gotz_decisionflow:_2014}.  The paper concludes with a discussion of
limitations and outlines key areas for future work.

%% file: background.tex
\section{Related Work}
\label{sec:background}

The IR approach to visual model validation is informed by advances in several
different areas of research.  These include the topics of uncertainty,
predictive visualization, and progressive or incremental visualization.  Also
relevant are visualization systems that utilize inferential statistics methods
and conceptual models of the visualization pipeline.

\subsection{Visualization of Uncertainty}

The visualization of uncertainty has been an active research area within the
visualization community for many years.  Studies have explored the problem
from many perspectives, including taxonomies that have
examined types of uncertainty \cite{skeels_revealing_2010} as well as
visualization methods for conveying uncertainty
\cite{potter_quantification_2012}.  In addition, there have been many efforts to 
formally study alternative methods for depicting uncertainty
measures~\cite{maceachren_visual_2012,sanyal_user_2009-1,tak_perception_2013}
through user studies that explore the perceptual understanding of
various uncertainty representations.  However, these studies focus on the visual
representation rather than methods for determining the degree of uncertainty.

Perhaps most relevant to the IR approach proposed in this paper is work that
has focused on estimating uncertainty via measures of entropy within a dataset
rather than by using carefully constructed statistical models
\cite{potter_visualization_2013}.  Like IR, this work adopts 
a non-parametric approach which does not require formal modeling nor make
assumptions about specific distributions within the data.
 
Finally, the distinction between the ``visualization of uncertainty'' and ``the
uncertainty of visualization'' has been highlighted \cite{brodlie_review_2012}.
The latter is a related but separate concept from traditional uncertainty
visualization.  This work highlights that the rendered graphics of a
visualization can convey a sense of authority which may not be warranted, even
when the underlying data itself is considered to be beyond reproach.  This
challenge is a key motivation for IR, as outlined in the discussion presented in
Section~\ref{sec:prediction}.

\subsection{Predictive Visual Analytics}

Visualization has long been used to support predictive analysis tasks.
However, most often, the ``prediction'' is performed by users reviewing historical
data and making assumptions about what might happen in the future for similar
situations.  In fact, the relatively limited history of work on visualizations
that incorporate more formal predictive modeling methods was the topic for a
workshop at the most recent IEEE VIS Conference in 2014 \cite{perer_ieee_2014}.

The work that does exist in this area has often focused on model
development and evaluation rather than supporting end users' predictive
analysis tasks.  For example, BaobabView \cite{van_den_elzen_baobabview:_2011}
supported interactive construction and evaluation of decision trees.  More
recent work has focused on building and evaluating regression models
\cite{muhlbacher_partition-based_2013}.  This method, like ours, adopts a
partition-based approach to avoid making structural assumptions about the data.
However, the focus on building regression models leads to an overall workflow
that is very different from the proposed IR approach.

Others have focused on visualizing the output produced by predictive models.
For example, Gosink et al. have visualized prediction uncertainty based on
formalized ensembles of multiple predictors \cite{gosink_characterizing_2013}.
This approach, however, requires careful modeling to develop the predictors, including 
the specification of priors that enable the Bayesian method that
they propose.  

Outside the visualization literature, where novel visual or interaction methods
are not a concern, predictive features are typically visualized using
traditional statistical graphics, for example, systems that visually prioritize
and threshold p-values to rank features for prediction (e.g.,
\cite{sipes_predictive_2011}).  Such methods are fully compatible with the IR
process proposed in this paper.

\subsection{Progressive/Incremental Visualization}

Overfit models and other sampling challenges are common to ``Big Data''
visualizations that rely on progressive or incremental techniques
(e.g., \cite{fisher_exploratory_2012,stolper_progressive_2014}).  Initial
samples are small, grow over time, and can change in distribution as time
proceeds.  Some have addressed this challenge by including confidence intervals
along with partial sets of query results \cite{fisher_trust_2012}.  However,
relying on the query platform to assess confidence in data subsets does not
easily support interactive zoom and filter operations after the query, because
these changes in visual focus do not necessarily result in new queries that
generate new result sets.  Moreover, these papers do not propose methods for
computing confidence intervals, but rather, assume that such data will be
provided by the the database.

\subsection{Inferential Statistics}

Statistical inference is a discipline with a very long and distinguished history.  
Most relevant to the IR method described in this paper are challenges related
to statistical significance and null hypotheses, and in particular Type~1 and
Type~2 errors.  Type~1 errors refer to improper rejections of the null
hypothesis which lead to conclusions that are not real effects, while Type~2
errors refer to falsely retaining the null hypothesis which can lead to
assumptions that a true effect is false~\cite{sheskin_handbook_2003}.  

These types of errors are of critical concern
in high-dimensional exploratory visualization where computational
methods can quickly access vast numbers of dimensions for statistical
significance.  Statistical correction methods have been proposed to reduce Type~1
errors \cite{stoline_status_1981}, but arguments have also been made against
this approach.  Those arguments suggest that parameterized models or
assumptions of ``default'' null hypotheses don't match real world situations where
distributions are rarely straightforward or independent.  Suggesting that these correction
methods are the wrong approach for exploratory work, Rothman argues that
``scientists should not be so reluctant to explore leads that may turn out to be
wrong that they penalize themselves by missing possibly important findings''
\cite{rothman_no_1990}.  

This tension is present in many interactive exploratory systems which make it
easy to generate vast numbers of potential hypotheses.  As a result, many
methods have been proposed for modeling measures of confidence or significance
\cite{chen_uncertainty-aware_2015,feng_matching_2010,wu_visualizing_2012}.
These efforts, however, typically rely formal statistical methods that make assumptions
about distributions and variable independence.

This approach is problematic for exploratory visualizations which allow users to quickly apply filters or
constraints that can quickly change the underlying assumptions.   The IR method we
propose provides provides a way for users to visually assess the
reliability of hypotheses.  Similar approaches that rely on user judgement have
been shown to be quite effective \cite{majumder_validation_2013}.

\subsection{Models of the Visualization Pipeline}

The traditional visualization pipeline model describes the process of transforming raw data
to an analytical abstraction, to visualization abstraction, and then finally to a
rendered graphic for interaction \cite{card_readings_1999,chi_taxonomy_2000}.
We add partitioning and aggregation stages to this flow to support the IR
approach.  As we will describe, a special case of the IR model (in which we
generate just one partition) is equivalent to the traditional model.  By extending the canonical pipeline, our work has similarities with Correa et al.'s paper describing pipeline extensions for an uncertainty framework focused on the data transformation process
\cite{correa_framework_2009}.

%% file: prediction.tex
\section{Visualization as a Predictive Model}
\label{sec:prediction}

Visualization design is often conceptualized as a mechanism for reporting.
This retrospective approach is so ubiquitous that terms such as
\emph{prediction}, \emph{forecast}, and \emph{inference} cannot be found within
the indices of many leading visualization texts from the past 25 years (e.g.,
\cite{munzner_visualization_2014,tufte_envisioning_1990,ware_information_2004}).

\begin{figure*}[t]
  \centering
  \includegraphics[width=18cm]{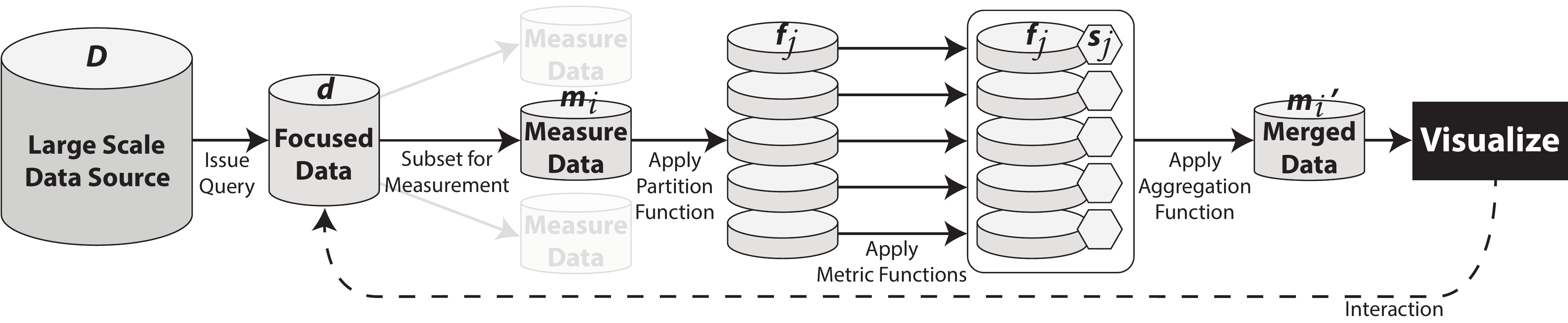}
  \caption{The Inline Replication (IR) visualization pipeline sends each
	  derived measure's subset of data ($m_i$) through a partition function
	  to create multiple folds ($f_j$) prior to mapping and visualization.
	  A metric function is applied to each fold independently, and an
  aggregation function recombines the folds to form an aggregate measure
  ($m_i'$) for subsequent visualization and interaction.}
  \label{fig:overall_flow}
\end{figure*}

Many visualization \emph{consumers}, however, use graphical representations of
historical data as the basis for decisions about future performance.  This is
done even when the underlying data and transformations do not support such
prospective conclusions.  Despite potentially fatal flaws
in terms of generalizability and repeatability, retrospective
visualizations are in essence being used as predictive models. 

\begin{figure}[t]
  \centering
  \includegraphics[width=8.2cm]{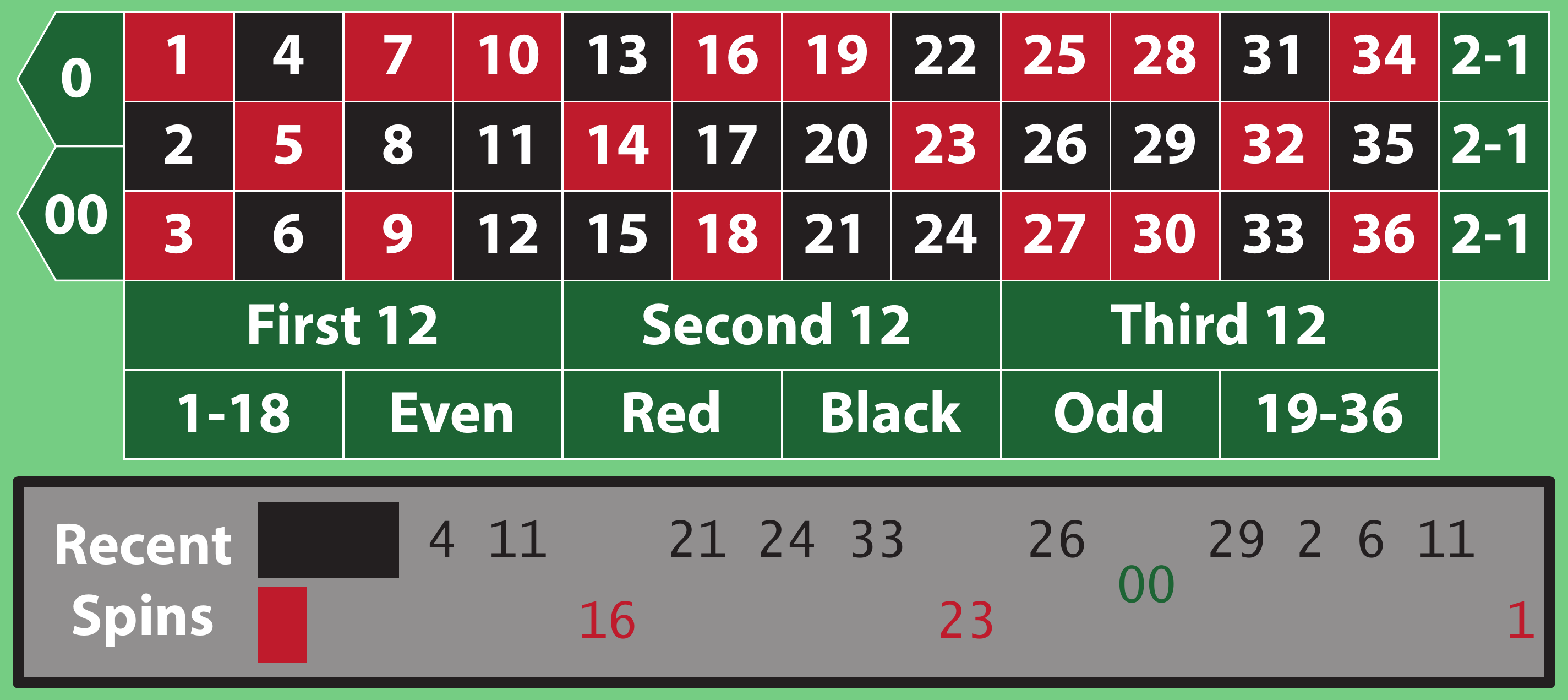}
  \caption{Roulette wheels allow users to bet on ``black'' or ``red'' squares.
  Casinos often display a simple visualization of ``recent spins'' to provide gamblers
  with a false sense of predictive knowledge.  In this example, the display
  shows a recent preponderance of black numbers with the implication to
  gamblers that this may influence future spins of the wheel.}
  \label{fig:roulette}
  \vspace{-0.2cm}
\end{figure}

The tendency to assume predictive power in visualization can be seen, for
example, in modern casinos.  Roulette wheels, for instance, commonly include an
electronic display (e.g., \cite{roulette_display_firm}) which shows the table's
recent history.  Assuming a fair table, ``red'' and ``black'' numbers should be
equally likely to appear.  However, as illustrated in Figure
\ref{fig:roulette}, the history provided to gamblers is not sufficiently long
to learn if the table has any systemic bias.  

Why then is the gambler presented with a simple visualization of the history?
The data is visualized to provide gamblers with a false sense of knowledge; to
suggest to a hesitant gambler that a bet is an informed decision rather than a
random choice.  A gambler may infer that the recent streak of black suggests
more black spins.  Alternatively, the gambler may infer an imminent return to
red.  To the casino, it does not matter what predictive inference is drawn as
long as it provides a sense of confidence that leads to increased betting.

It is tempting to dismiss this scenario as one in which the gambler should be
more informed about basic statistics.  The small sample size and the
independence of each roulette spin should make it clear that the display is not
especially informative.  However, relatively sophisticated users performing
visual analysis of data from more complex underlying systems can make similarly
poor predictive assessments on the basis of visual representations that don't
properly convey the underlying limits of their predictive power.  

For example, consider a business analyst attempting to learn about why sales
are declining, or a physician using historical patient data to compare treatment
efficacy.  In these complex real-world cases, in which it is essentially
impossible to fully understand the underlying statistical processes, it is
natural for analysts to turn to visualization as a predictive model for their
problem.  Visualization allows these users to see what has happened and, based
on trends or patterns in the representation, make assumptions about what will
happen in the future.  

However, just as the casino gambler draws inference from a not-so-meaningful
visualization, these power users can be led to make poor predictions on the
basis of visualizations that are essentially ``overfit'' models based on poor
representations of the underlying process.  This problem has even been
documented even in highly quantitative fields such as epidemiology, where
public health analysts have had trouble discounting statistics from small
sample sizes when visualized \cite{sutcliffe_developing_2014}.

Issues of poor sampling and overfit are especially problematic during
exploratory visualization in which users can interactively apply arbitrary
combinations of filters to produce new ad hoc subsets of data for
visualization.  Such systems are at greater risk of generating
non-representative visualizations that occur ``by chance'' rather than due to
real properties of the underlying problem.  The same is true for visualization
systems that utilize sampled or progressive queries to address issues of scale.

The potential for this sort of ``visual model overfitting'' is
analogous to the overfitting problem in more traditional modeling tasks.
In the machine learning community, this is addressed in
part by cross-validation, a widely used technique for assessing the
quality and generalizability of a model \cite{kohavi_study_1995}.  Rather than
relying on a single model, cross-validation methods create and compare multiple
models, one for each of several partitions of a dataset (often called
``folds'').  This allows for an assessment of model repeatability, with models
that work consistently across partitions considered more trustworthy.

If one considers---as we argue here---that a visualization is often used as a
form of predictive model, then validation becomes a critical guard against
problems associated with overfitting.  When a visualization is zoomed and
filtered to focus on a specific subset, is the visual representation
repeatable?  Are the conclusions drawn from the visualization generalizable?
Can an approach similar to cross-validation be embedded within the
visualization pipeline so that each new view produced during user interaction
is evaluated?  The IR method outlined in the next section is designed to
support this form of validation as an integrated part of the visualization
process.

%% file: theory.tex
\section{Inline Replication}
\label{sec:theory}

Inline Replication (IR) is an approach to visualization in which the dataset
associated with each visualized measurement is partitioned into multiple subsets (which
we call \emph{folds}), processed independently to calculate derived statistics
or metrics, then aggregated together to be mapped to a set of graphical marks 
and rendered. This partitioned approach
embeds an automated and non-parametric workflow for replication within the
visualization pipeline as illustrated in Figure~\ref{fig:overall_flow}. The
result is that visualizations based on IR provide users with important
information about the repeatability of observed visual trends, reducing the
likelihood of certain types of erroneous conclusions.  

The IR pipeline begins with the same initial step as a traditional
visualization pipeline.  As illustrated in Figure~\ref{fig:overall_flow}, a set
of query or filter constraints is applied to a primary data source $D$ to
produce a focused dataset $d \subset D$.  The data in $d$ is then organized
into subsets for which statistical measurements are calculated, creating
measure-specific subsets of data which we note as $m_i$.  For example, a
visualization pipeline configured to generate the bar chart in
Figure~\ref{fig:roulette}, showing the distribution between black and red spins
for a roulette wheel, would include the subset $m_{recent}$ containing 
data for the most recent spin results (black or red).  If the visualization included multiple bar charts (e.g., past 10 spins, past 100 spins, and past 1000 spins) then multiple subsets $m_i$ would be defined because each requires the calculation of a distinct set of measurements.

Traditionally, the data for each subgroup $m_i$ would immediately be processed
to compute the measurements required for visualization (e.g., the fraction of
spins resulting in black, and the fraction of spins resulting in red).  Those
measures would then be mapped to visual properties of objects within the
visualization (e.g., the size of each bar in the bar chart).

The IR pipeline, however, behaves differently.  Each $m_i$ is first partitioned
into distinct folds $f_i$, each of which is analyzed independently via a metric
function.  The results are then aggregated to form a merged dataset $m_i'$.  It
is this merged representation of the measures, $m_i'$, that is mapped to the
visual representation and rendered to the screen for interaction using methods
designed to convey the repeatability of the visual model across each of the
folds.

This section provides an overview of the IR pipeline, focusing on the three
functions at the core of the design: the partition function, the metric
function, and the aggregation function.  It then describes the IR approach to
visual display and interaction, and concludes with a discussion of useful
variations to the core design.

\subsection{Partition Function}
\label{sec:partition}

Conceptually, the \emph{partition function} is designed to subdivide the data
in a given measure-specific subset $m_i$ into multiple partitions.  The goal of
this stage in the IR pipeline is the creation of several independent datasets,
which we call \emph{folds}, to use as the basis for calculating each
measurement.  Later in the IR process, derived measures (e.g., proportions, or
statistical significance) will be calculated for each fold.

Formally, we define the $Partition$ function as an operator that subdivides a
measure-specific set of data $m_i$ into $n$ folds such that
each fold $f_j \subset m_i$.

\begin{equation}
Partition(m_i, n) \rightarrow \{f_0, f_1, \cdots, f_j, \cdots, f_n\}
\label{eq:partition}
\end{equation}

This function is applied to the raw data in $m_i$, prior to any other
aggregating transformations (such as the summation in the roulette example).
Following an approach inspired by $k$-fold
cross-validation~\cite{kohavi_study_1995}, the baseline partition function
creates $n$ folds that are disjoint, approximately equal in size, and randomly
partitioned such that:

\begin{equation}
\bigcup_{j=0}^{n}f_j=m_i
\label{eq:comprehensive}
\end{equation}

As discussed previously, multiple folds are created with the goal of supporting
repeated calculations for each measure.  Increasing the value of $n$ to produce
more folds increases the replication factor.  However, higher $n$ values also
produce smaller $f_j$.  If $n$ is too large for given $m_i$, the folds may be
too small to compute useful measures.  Therefore, $n$ can be dynamically
determined so as to require a minimum fold size.  If $m_i$ represents a
``large enough'' subset of data, it will produce a full set of folds.  If,
however, $m_i$ is too small for the minimum fold size, fewer than $n$ folds
will be produced.  The threshold for ``large enough'' depends on many factors,
including the specific metrics that will be calculated.

Partitioning with $n=1$ results in the \emph{identity partition function} where
$f_0 = m_i$ regardless of the size of $m_i$.  Because no partitioning is
performed, an IR process using the identity partition function produces results
that are identical to a traditional visualization pipeline: a single metric is
calculated and visualized.  In this way, the traditional approach to
visualization can be seen as a special case of the IR process in which
replication is not performed because there is only one fold.

Choosing a proper $n$ value is necessarily a compromise between increased
replication and smaller sample size.  We can look to the machine learning
community for guidance, however, where empirical studies have shown that there
is no meaningful benefit for values of $n$ over 10~\cite{kohavi_study_1995}.
Moreover, as datasets grow larger in many fields, smaller samples become less
of a concern.

Finally, there are certain conditions (e.g., very small datasets with little data
to partition, or very large datasets where sampled queries are required) where
the basic formulation for the partition function can be problematic.
Variations to the partitioning process, designed to help address these 
challenges of scale, are discussed in Section~\ref{sec:variations}. 

\begin{figure*}[t]
  \centering
  \includegraphics[width=18cm]{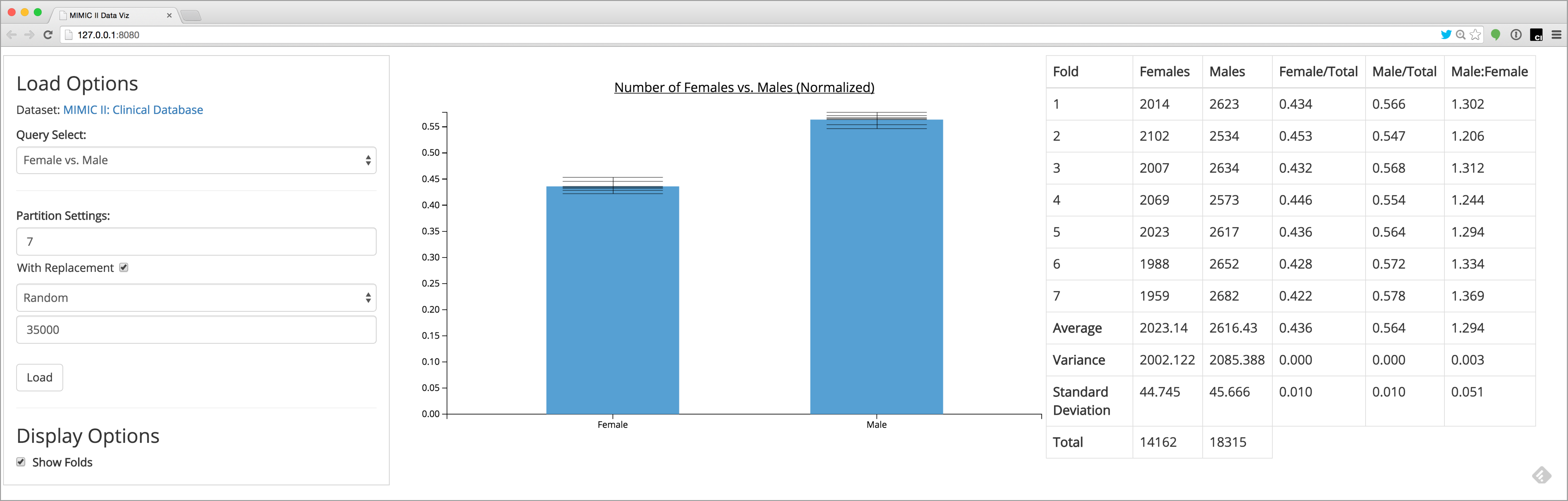}
  \vspace{-0.05cm}
  \caption{The IR-based prototype shown here was developed to test the proposed
  pipeline and to explore the parameter space with two baseline visualization
  types: bar charts and linear regression lines.  The left panel shows the
  query and IR controls, the middle panel shows the visualization space, 
  and the left panel shows detailed descriptive statistics computed for both the
  aggregate representation and the individual folds.}
  \label{fig:baseline_prototype}
  \vspace{-0.1cm}
\end{figure*}

To illustrate the partitioning process, consider the roulette example from
earlier in this paper.  The example bar chart showing the fraction of spins
resulting in black or red is based on a single measure-specific subset of
data $m_{recent}$. The $Partition$ function would be applied to this subset to
create a set of multiple folds, each of which would contain a subset of the
recent spin results.  For example, $Partition(m_{recent}, 5)$ would produce a
set of five folds, each containing results from one-fifth of the overall set of
recent spins.

\subsection{Metric Function}

The folds produced during partitioning are sent to a \emph{metric
function} which is applied independently to each fold as illustrated in
Figure~\ref{fig:overall_flow}.  The goal of the metric function is
to derive a set $s_j$ of one or more derived statistics for each fold $f_j$.
Because the metric function is applied to all folds, multiple sets of
statistics are created for each $m_i$.  These statistics can then be aggregated
and compared during the eventual visualization rendering process.  

The specific measures computed by the metric function are necessarily
application specific, but could range from simple descriptive statistics
(e.g., sums, averages) to more complex analyses (e.g., classification,
regression).  Generally speaking, the metric function is defined to produce the
same derived values that would normally be computed as part of a more
traditional visualization process.  The key difference in IR is
that the metrics are computed multiple times for $m_i$ (once for each
fold), where traditionally such values would be computed just once.

For example, consider the roulette use case described earlier.  The metric
function in this example would compute the fraction of spins resulting in black
and red in each fold $f_j$.  This fraction is the same measure that the
original bar chart is designed to display.  However, with the IR approach, the
metric is calculated for each of the five folds produced by
$Partion(m_{black},5)$. 

An actual implementation of IR using a similar ``fraction of the population''
metric function is discussed in Section~\ref{sec:usecase}.  However, more
sophisticated systems may adopt more advanced measures.  For example,
correlation statistics, p-values, metrics of model ``fit'', and regression lines
are all compatible with the IR approach.  Examples of IR using linear
regression, correlation, and statistical significance testing are all described
in Section~\ref{sec:usecase}.

\subsection{Aggregation Function}

The metric function produces a set of statistical measures $s_j$ for each of
the $n$ folds $f_j$ that are produced by the partition function.  Prior to
visualization, the multiple $s_j$ metrics must be aggregated to a single
representation $m_i'$ to invert the partition process.  As illustrated in
Figure~\ref{fig:overall_flow}, this is accomplished via an \emph{aggregation
function} which we define as follows.

\begin{equation}
Aggregate(\{(f_j, s_j)\}) \rightarrow m_i'
\end{equation}

The \emph{Aggregate} function is designed to produce one aggregate value for
each of the different measures computed by the metric function.  For instance,
if a metric function computes two measures for each fold (e.g., count and
correlation), then the aggregation function would produce two corresponding
aggregate measures.

A variety of aggregation algorithms can be employed, with different approaches
appropriate to different types of metrics.  For example, for count-based
metrics which capture the frequency of data items in each fold, a summation
across all folds might be the most appropriate because a sum of counts for each
fold provides an accurate total for the overall data subset $m_i$.  For a
metric that captures a mean or rate, averaging the values across all folds may be 
most appropriate.  For categorical metrics, meanwhile, such as those produced
by classification algorithms, a ``majority vote'' aggregation method
\cite{lam_application_1997} can be applied to capture the most frequently
assigned category.  The same voting approach can be used when aggregating thresholded measures (e.g., tests of statistical significance) across all folds.  This approach is demonstrated in Section~\ref{sec:decisionflow_usecase}.

The summary measures produced by the aggregation function are combined with the
set of $s_j$ statistics computed for the $n$ folds to form a merged data
representation $m_i'$.  This merged representation is then used to drive the
mapping and rendering process of the final visualization.

As a concrete example, consider again the roulette scenario.  The metric
function described previously computed the fraction of spins resulting in black
and red numbers for each of the five folds created by $Partition(m_{recent},
5)$.  Because the partitions are by definition equal in size, aggregate rates
for both colors can be obtained by averaging the five fold-specific rates.  The
overall average, along with the individual values for each fold, are combined
to form $m_{recent}'$.

\subsection{Visual Display And Unfolding of Partition Data}

Once aggregation has been performed, the merged data $m_i'$ is mapped to its
corresponding visual marks and displayed as part of the visualization.  This is
shown as the final step in Figure~\ref{fig:overall_flow}.  The IR approach to
visualizing $m_i'$ has two elements, which correspond to the two distinct types
of information in the merged data structure: (a) the aggregate measures and (b)
the individual fold measures.

First, an initial visualization is created using only the aggregate measures.
The process for this stage is similar to a traditional visualization pipeline.
The aggregate measures are mapped to visual properties of the corresponding
graphical marks, which we call \emph{aggregate marks}.  These marks
are then rendered to the screen for display and interaction.
In the roulette example, for instance, the aggregate data for black and red
spin rates (produced by the $Aggregate$ function) can be used to generate a basic bar chart that is identical to what is shown in Figure~\ref{fig:roulette}.

\begin{figure*}[t]
  \centering
  \includegraphics[width=18cm]{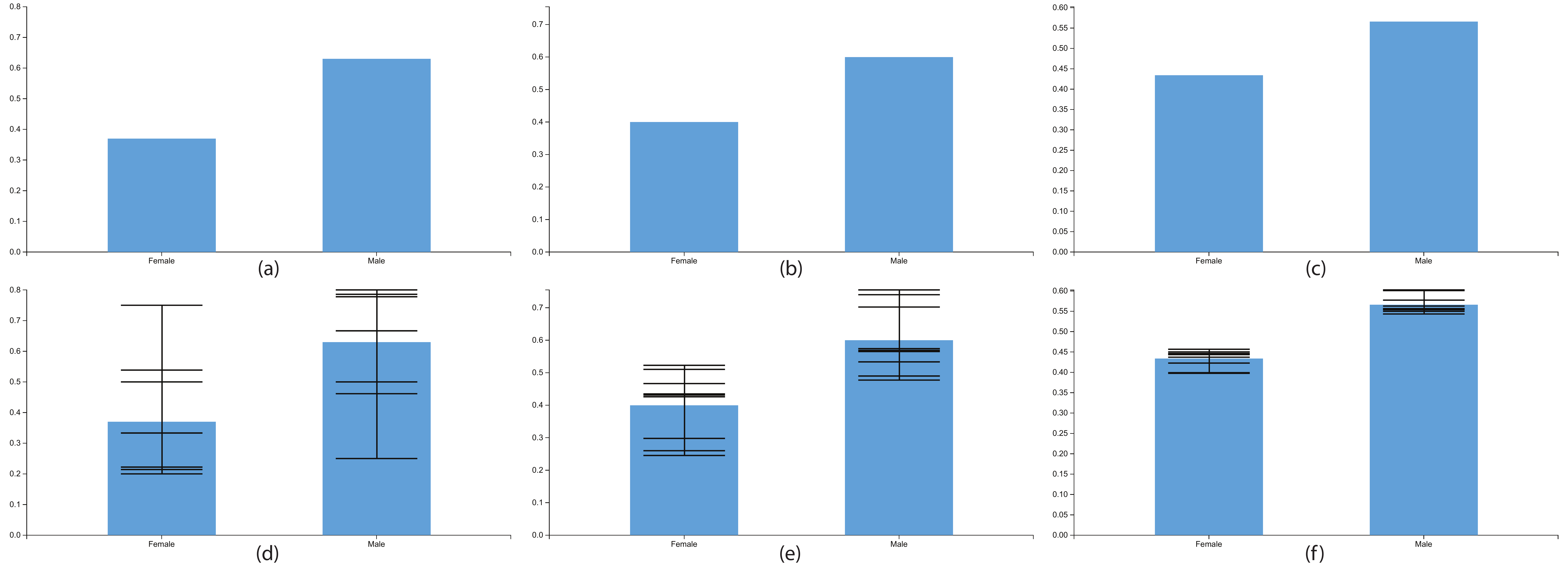}
  \caption{Six charts produced by the IR prototype system.  The top three charts (a-c) show the gender distribution for three
	  different sets of ICU patients.  The relatively similar bar charts
	  suggest that the underlying populations are comparable.  However, when
	  the same populations are visualized with 7 folds
	  (d-f), a different story appears.  The charts now clearly demonstrate 
	  that we know less about the population visualized in the left column
	  than we do about the population on the right.  In this
case, the difference is due largely to the size of the respective populations.}
  \label{fig:confidence}
\end{figure*}

Second, an IR visualization allows aggregate marks to be \emph{unfolded}.  An
unfolding operation---typically triggered by a user interaction event such as
selection or brushing---augments the aggregate marks with a visualization of
the individual fold statistics that contribute to the aggregate measures.  In
the ongoing roulette example, the fold data would show the variation in
proportion of spins that result in black and red numbers across each of the
$n=5$ independent folds.  Additional examples from our experimental prototypes
are described in Section~\ref{sec:usecase}.

\subsection{Discussion}

The ability to unfold aggregate measures into repeated measurements 
is a central contribution of the IR approach.  By
graphically depicting the repeatability of a particular measure across
multiple folds, IR provides users with important and easy-to-interpret cues as
to the variability of a given measure.  Traditional visualization methods do
not convey this information, meaning it is often not considered when
predictive conclusions are made by users.

Another benefit of IR comes from the aggregation function.  In particular,
embedding within the visualization pipeline an ability to aggregate
categorical values such as statistical significance classification can lead to
more accurate results.  Repeated measures combined with voting-based
aggregation can, for instance, reduce the exposure to Type 1 errors when
looking for statistically significant p-values.  For example, a statistically
significant ($p<0.05$) run of black spins on the roulette wheel is less likely
to occur ``by chance'' across a majority of $n$ folds than it is across a
single group of spins.  This is a major benefit for exploratory visualization
techniques that allow users to visually ``mine'' through large numbers of
variables in search for meaningful correlations.

\subsection{Variations}
\label{sec:variations}

Following the traditional approach to $k$-fold cross-validation, the baseline
$Partition$ function defined in Section~\ref{sec:partition} specifies that the
constructed folds are disjoint, randomly partitioned, and exhaustive (Equation~\ref{eq:comprehensive}).  However, relaxing these constraints leads to several valuable variations to the baseline IR procedure.

{\bf Partial Partitioning.}  Relaxing the requirement of
Equation~\ref{eq:comprehensive} allows for the creation of partitions that do not
contain all data points within $m_i$.  For very large datasets, this can allow
for approximate analyses that use only a subset of the available data.  This
approach provides significant performance benefits for metric functions that
have poor scaling properties, and it allows IR to work directly with recently
proposed techniques for progressive visualization (e.g.,
\cite{fisher_exploratory_2012,stolper_progressive_2014}).

{\bf Partitioning With Replacement.} Relaxing the requirement that all folds
are disjoint allows for partitioning with replacement.  Similar to statistical bootstrapping, this approach allows the same data point to be included in multiple
folds (or even multiple times within the same fold).  When allowing replacement,
the dataset in $m_i$ becomes a sample distribution from which the partitioning
algorithm can generate a larger population. 
This is especially useful for small datasets---a frequent occurrence in
exploratory visualization where multiple filters can be quickly
applied---because the larger generated population can allow the IR process to
run with less concern about producing fold sizes that are too small.

{\bf Incremental Partitioning.} A number of progressive or sampled methods have
been proposed in recent years to address the challenges of ``Big Data''
visualization (e.g., \cite{fisher_trust_2012,stolper_progressive_2014}).  In
these approaches, the full dataset $m_i$ is often never fully retrieved.  To
utilize an IR approach in these cases, an incremental partitioning process is
needed.  During this process, data points should be distributed to folds as
they are retrieved such that all $n$ folds are kept roughly equal in size.
This will allow IR to work with continuing improvement in metric quality as
more data arrives.  However, we note that IR will not overcome selection
bias that may be introduced as part of the progressive query process.  Therefore,
the determination of a progressive sampling order that is both representative
and balanced remains a critical concern.

%% file: usecase.tex
\section{Use Cases}
\label{sec:usecase}

The IR approach is compatible with a broad range of visual metaphors and
interaction models, from basic charts to more sophisticated exploratory visual analysis
systems.  To demonstrate this flexibility and to explore the impact of adopting an IR
pipeline, we developed two prototype IR systems: (a) a reference prototype to
study IR in isolation, and (b) a more sophisticated visual analysis system to
examine IR within a more complex analysis environment.

\subsection{Prototype 1: Reference Prototype}
\label{sec:baseline_usecase}

We developed a reference IR implementation as part of a simplified visual
analysis prototype with the goal of exploring the IR parameter space in
isolation, without concern for the more complex interactions that are part of
a real-world application such as the one described in Section~\ref{sec:decisionflow_usecase}.  The prototype supports
two basic visualization metaphors: (a) bar charts and (b) scatter plots with
linear regression lines.  The prototype was tested using a dataset of
electronic medical data containing over 40,000 intensive care unit (ICU) stays
\cite{goldberger_physiobank_2000}.

\begin{figure*}[t]
  \centering
  \includegraphics[width=18cm]{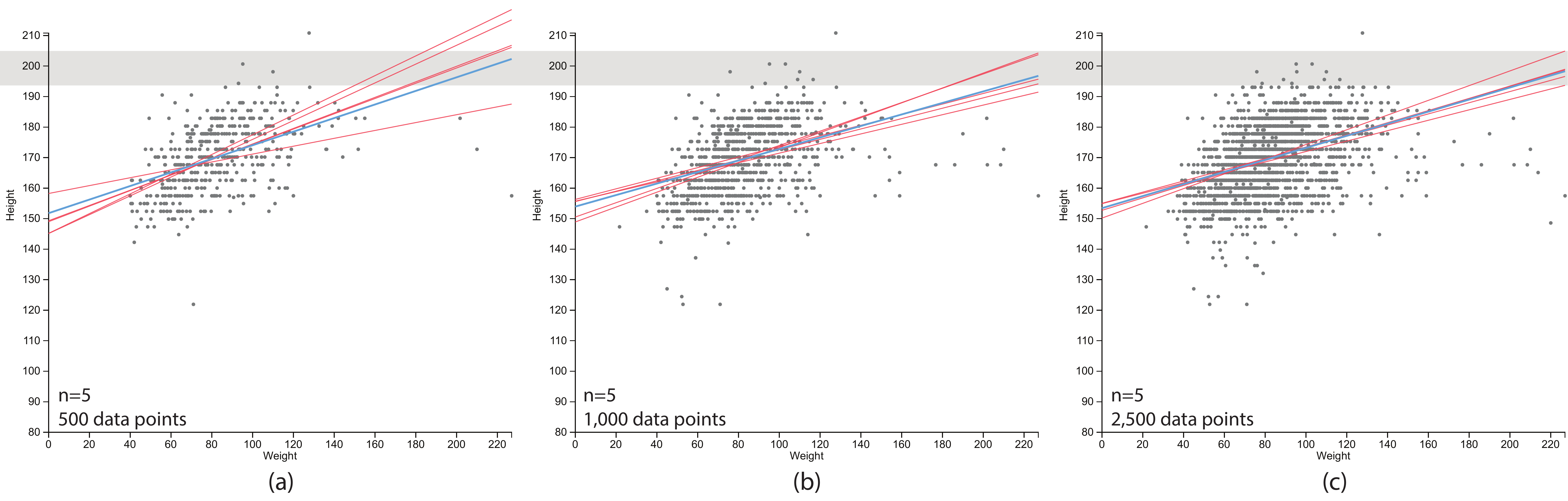}
  \vspace{-0.1cm}
  \caption{Weight versus height distribution for patients admitted to a
  neonatal intensive care unit.  Simulating the results from a progressive
  visualization system, this figure shows both the raw data and best fit
  regression line (shown in blue) for (a) 500 patients, (b) 1,000
patients, and (c) 2,500 patients.  In all three cases, the IR pipeline has
computed a regression across five folds, shown in red.  The decreasing
spread across the red regression lines conveys the expected---but often
overlooked---change in variation between folds as the sample size
increases. The gray band across all three charts has been added to this figure
to emphasize these differences and reflects the variation across folds in (c)
at the maximum observed weight.}
  \label{fig:weight_regression}
  \vspace{-0.15cm}
\end{figure*}

The prototype interface, shown in Figure~\ref{fig:baseline_prototype}, includes
three  panels.  In the center is the visualization canvas itself.  A left-side
panel allows users to issue queries and control key parameters to the IR
process.  Options include the number of folds ($n$) for the partition function,
the use of sampling with replacement, support for random or ordered incremental
sampling, and controls to unfold the merged statistics to show individual folds
within the visualizations.  The right-side panel shows detailed descriptive
statistics for both the individual folds ($s_j$) and the aggregate dataset
($m_i'$).

Figure~\ref{fig:confidence} shows a series of bar charts rendered
using the IR prototype to visualize the gender distribution across three
subpopulations from the ICU stay database.  This example is directly analogous
to the roulette wheel bar chart example introduced in
Section~\ref{sec:prediction}, as both summarize the distribution of a binary
variable in a given population.  

The top row of charts in Figure~\ref{fig:confidence} shows the aggregate gender distribution
for each of the three populations.  The charts show a relatively similar
distribution across all three populations, with a moderate increase in female
representation moving from panel (a) to (b) to (c).  The bar chart shows the
gender breakdown in each population quite clearly.  However, there is no
indication of the distribution's stability across different groups of patients.
Consumers of the visualization are left to assume that the bar charts provide
an accurate depiction.

Panels (d-f) in Figure~\ref{fig:confidence} show the exact same populations
as panels (a-c), respectively.  However, these views incorporate measures
computed for multiple folds ($n=7$) using the IR process of partitioning and
merging.  These unfolded views provide a more accurate picture regarding the
repeatability of the gender distributions in the top row of the figure.  In
particular, we see from Figure~\ref{fig:confidence}(d) that the population
visualized in the left column of the figure is not very predictable.
Meanwhile, far less variation across folds is visible in
Figure~\ref{fig:confidence}(f).  In this case, the major difference is the size
of the respective populations which range from about 100 to roughly 10,000
patients.  This is a critical factor to interpretation which is invisible in
the original bar charts and easily overlooked even by expert users (e.g.,
\cite{sutcliffe_developing_2014}).  

Figure~\ref{fig:weight_regression}, meanwhile, shows three screenshots of the
linear regression portion of the IR prototype applied to data from the same ICU
repository used for the bar charts. In this case, the examples show data for
populations of neonates on a scatter plot, with the x position determined by
weight and the y position determined by height.  A linear regression model was
calculated in all three cases using the IR pipeline with $n=5$.  The five
regression lines, one for each fold, are visible (``unfolded'') in the
visualizations as red lines.  In addition, an aggregate best-fit linear model
is shown in blue.

To explore how IR helps convey uncertainty during progressive analysis, we
used the incremental sampling feature of the prototype to vary the number of
samples while keeping all IR parameters constant.  In
Figure~\ref{fig:weight_regression}(a), only 500 patients are included in the
scatter plot.  As the varying slopes between the five red lines captures, there
is relatively large disagreement across folds in the linear models they
produce.  This uncertainty would be invisible in a traditional plot rendered without
the folds.  

As expected, the spread between the individual fold regression lines decreases
as more patients are retrieved by the incremental query feature.  For example,
Figure~\ref{fig:weight_regression}(b) shows the same visualization with the
same $n=5$ folds.  However, this version includes data for 1,000 patients.  The
larger sample size results in increased stability across the folds.  Part (c)
of the same figure shows the same visualization with 2,500 patients.  We see
little improvement in agreement across folds compared to 1,000 patients,
suggesting that the rate of further gains in agreement will be slower to develop.

As previously stated, the improvement in agreement as sample size increases is
as expected.  However, as evidenced by the ``recent history'' charts at casino
roulette tables and the other examples referenced throughout this paper,
visualizations are often assumed to be accurate.  Users often fail to consider
issues of sample size or variation.  This use case shows that IR can
effectively convey this variation in the data without careful modeling, and in
a non-parametric way that avoids assumptions about the underlying
distributions.

\subsection{DecisionFlow2}
\label{sec:decisionflow_usecase}

To test IR within a more fully-featured exploratory visual analysis
environment, we developed DecisionFlow2, a new IR-based version of our existing
visual analysis system for high-dimensional temporal event sequence
data \cite{gotz_decisionflow:_2014}.  A screen capture of the DecisionFlow2 interface
is shown in Figure~\ref{fig:teaser}.

\subsubsection{Original DecisionFlow Design}

The original version of DecisionFlow made heavy use of
p-values to help users identify event types that had a statistically
significant correlation to a user-specified outcome measure.  When visualizing
medical data, for example, this approach allows users to find types of medical events
(such as specific diagnoses, medications, and procedures) that---when appearing in a
particular pattern in a patient's history---are associated with better or worse
medical outcomes.  

In the original DecisionFlow design, an interactive timeline at the top of the
screen allows users to segment a cohort of event sequences based on the
presence of so-called ``milestone'' events.  For a given subgroup, DecisionFlow
visualizes statistics for the potentially thousands of different types that
occur between milestones with the goal of helping users identify good
candidates for new milestones.  DecisionFlow conveys the event type statistics
via an interactive bubble chart similar to the one seen in Figure~\ref{fig:teaser}.

In the bubble chart, each event type is represented by a circle whose x-axis
position is determined by its positive support (the fraction of ``good
outcome'' event sequences that contain the event type).  Similarly, each
circle's y-axis position is determined by its negative support (the fraction of
``bad outcome'' sequences with the event type).  Circle size and color encode
correlation and odds ratio, respectively.  Importantly, circles representing
event types whose presence correlates significantly ($p<0.05$) with outcome are
drawn with a distinct border to make it easier for users to visually
distinguish between expected variation and potentially meaningful associations.

\subsubsection{Design Adaptation for IR}

In the IR-based DecisionFlow2 system developed for this paper, a similar
bubble chart design design is adopted to visualize the event type statistics.
However, rather than showing data for measures computed for the overall
population $m_i$, the circles encode aggregate measures computed by an
aggregation function.  For example, Figure~\ref{fig:teaser} shows the system
with a bubble chart focused on subset of data containing 45,278 individual
events with 1,148 distinct event types.  The support values (used to position
the circles) and other measures were all computed across 5 folds.

The aggregate view in Figure~\ref{fig:teaser}(a) looks essentially identical to the original
DecisionFlow design.  This is as intended, with the goal of making IR
compatible with typical visualization designs.  However, while the visual
encoding is similar, the number of statistically significant correlations
scores is reduced.  In particular, a number of event types that were labeled as
statistically significant in the original design were no longer found to be
significant once majority-voting across the five folds was used to determine
which event types were significant.  This makes the visualization system more
selective in rejecting the null hypothesis.  The result is a reduction in the
likelihood of Type~1 errors, which are a common problem in high-dimensional
exploratory analysis.  More detailed results and discussion are provided in
Section~\ref{sec:df_results}.

Another important part of the IR-based DecisionFlow2 is the ability to
unfold the aggregate statistics for each event type.  Users can unfold an event
type by hovering the mouse pointer over the corresponding circle.  For example,
after hovering the mouse pointer for a few seconds over the circle shown in
Figure~\ref{fig:teaser}(a), the unfolded representation shown in
Figure~\ref{fig:teaser}(b) is added to the visualization.  

As this example shows, the DecisionFlow2 displays the unfolded
data as a convex region drawn around the original circle and outlined with a dashed
border.  This region corresponds to the convex hull determined by the $(x,y)$
locations for each of the $n$ folds that contribute to the aggregate measures
that determine the position of the original circle.  In other words, the size
and position of the unfolded region represent the variation across folds in
both the positive and negative support measures.  Smaller unfolded regions
indicate that the values have little variation across folds.  Larger unfolded
regions, such as the one shown in Figure~\ref{fig:teaser}(b), suggest a high
degree of variation between folds and therefore lower confidence in the
repeatability of the aggregate measure.

\begin{figure}[t]
  \centering
  \includegraphics[width=9cm]{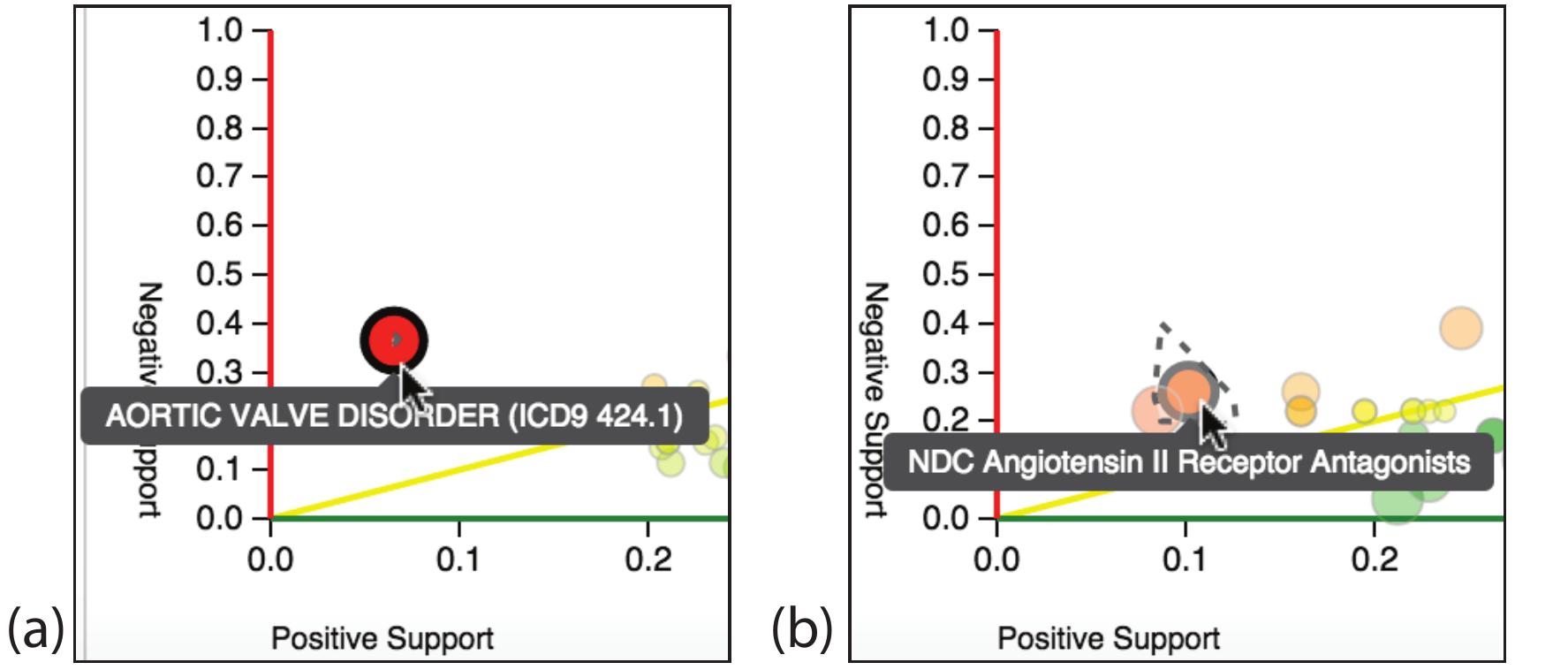}
  \caption{In general, (a) smaller differences between folds are seen
  when sample sizes are larger, while (b) higher levels of variation are seen
  for smaller sample sizes.}
  \label{fig:sample_sizes}
  \vspace{-0.14cm}
\end{figure}

The typical behavior observed when utilizing the IR-based implementation of
DecisionFlow2 is shown in Figure~\ref{fig:sample_sizes}.  Part (a) of the figure
shows an event type from a very large subset of data that shows very limited
variation across folds.  This is represented by the very small unfolded region
located near the center of the red circle just above the mouse pointer.   Part
(b) of the figure, meanwhile, shows an event type with much higher variation.
This figure, visualizing data from a smaller sample size, demonstrates
what one might expect: findings based on smaller sample sizes have more
variability and therefore should be given less weight in a decision making
process.

However, this very critical difference is not observable via the original
bubble chart.  The size of the dataset is made available elsewhere in the user
interface for users who consciously seek it out, but the implications of the
differences in data size are left to the user's imagination.  It is only
through the unfolding process that the visualization itself conveys the
difference in confidence that users should place in one view versus the other.

\begin{figure*}[t]
  \centering
  \includegraphics[width=18cm]{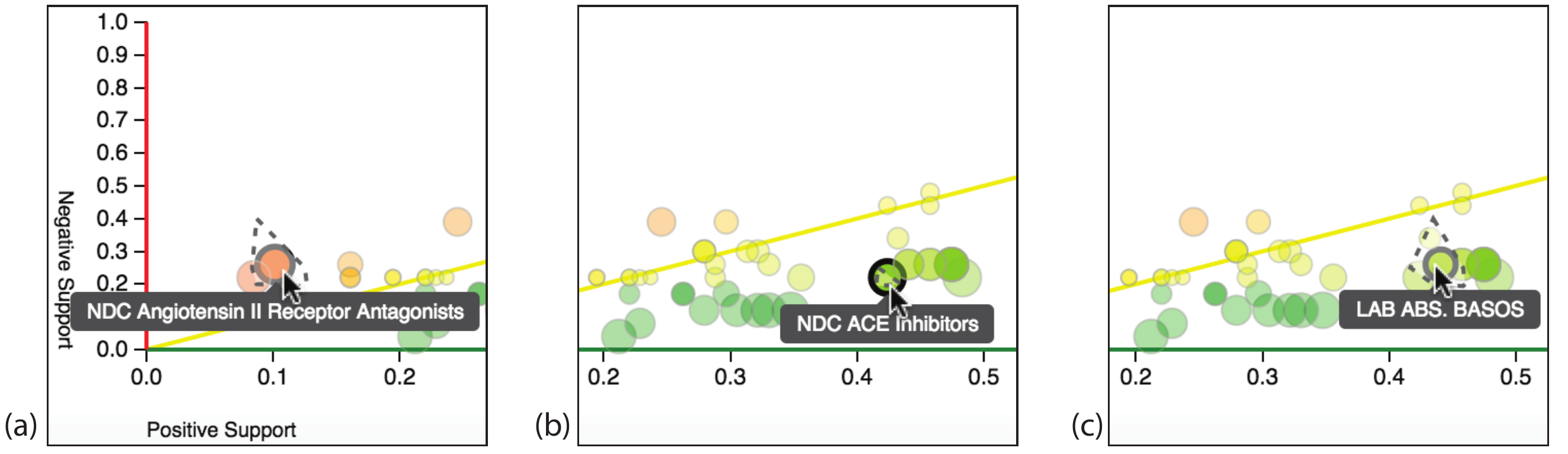}
  \caption{Even with the same sample size, different measures can have
  different levels of repeatability across folds. In this example, both (a) and
  (c) show relatively high levels of variability, while the small unfolded
  region in (b) suggests that the relationship between outcome and ACE
  Inhibitors was fairly consistent across all five folds.  All three views were
  calculated using identical sample sizes.}
  \label{fig:variation_same_sample_size}
  \vspace{-0.2cm}
\end{figure*}

Moreover, it is critical to note that the size of the dataset is not the sole
determinant of repeatability for a given measure across folds.  Major
differences in measure values can be seen even for similarly sized datasets.
For example, Figure~\ref{fig:variation_same_sample_size} shows three different
event types from the exact same subset of event sequences.  While the number of
event sequences was the same, the association between \emph{ACE Inhibitors} and
the user-defined outcome (eventual diagnosis with heart failure) was far more consistent across folds.

\subsubsection{Results and Analysis}
\label{sec:df_results}

The IR-based DecisionFlow2 prototype provides visual feedback regarding the
variation in positive and negative support.  As previously described, the
system also uses IR to assess the statistical significance of each event type's
correlation with patient outcome.  For a given event type, correlation
coefficients and p-values are computed for each fold, then aggregated via
majority-vote.  Event types with more than $n/2$ folds showing statistical
significant correlation are displayed in the visualization with a distinct
visual encoding.

To better understand the impact of IR and the choice of $n$ on the visualized
results, we conducted a quantitative experiment in which we compared performance
for a sample user interaction sequence under various conditions.  More
specifically, we experimented repeatedly by performing the exact same
exploratory analysis steps using DecisionFlow2, using the exact same input
data, varying only the number of folds.  The experiment was conducted at three
partition settings: $n=\{1,3,5\}$.  

In all three cases, the input dataset consisted of event data from the medical
records of 2,899 patients containing 1,074,435 individual medical events.
These timestamped events contained 3,631 distinct medical event types: specific
diagnoses, lab tests, or medication orders that were present in the patients'
records.  

Of the 3,631 distinct event types, 381 were deemed prevalent enough by the
DecisionFlow2 system to be the target of correlation analysis within the metric
function.  The same threshold was used across all three partition settings,
allowing us to compare analysis results across the exact same control
conditions.  The results of our analysis are shown in
Table~\ref{tab:df_results}.

With $n=1$, the DecisionFlow2 system flagged statistically significant results
in the same way as in the original paper \cite{gotz_decisionflow:_2014}.  Using
a threshold of $p<0.05$, 144 statistically significant event types were
detected.  When $n$ was increased to three, the numbers were reduced
dramatically.  Only 50 of the original 144 statistically significant event
types remained after applying a majority-vote aggregation algorithm.  Of those
50, only 43 were significant across all three folds.  For $n=5$, the number of
significant event types was even smaller.  The stricter requirements for
replication resulted in just 24 event types being flagged as significant given
a majority-vote aggregation algorithm, and just 15 event types were significant
across all 5 folds.

\begin{table}[t]
  \centering
  \vspace{0.3cm}
  \begin{tabular}{lccc}
    \toprule
    Number of Folds & $n=1$ & $n=3$ & $n=5$ \\
    \midrule
    Unanimously Significant      & 144 & 43 & 15 \\
    Majority Significant         & 144 & 50 & 24 \\
    At Least One Significant     & 144 & 56 & 29 \\
    \midrule
    Total Number of Measurements Made & \multicolumn{3}{c}{381} \\
    Total Number of Event Types  & \multicolumn{3}{c}{3,631} \\
    \bottomrule
  \end{tabular}
  \caption{A comparison of statistically significant findings in
  three different IR configurations with DecisionFlow2 applied to the same data.
  The number of event types flagged as significantly associated with outcome
  was largest for $n=1$.  This setting corresponds to a traditional visualization
  approach with no partitioning.  Larger $n$ values dramatically reduced the
  number of significant findings.}
  \label{tab:df_results}
\end{table}

As expected---and as intended---the number of statistically significant findings
is reduced as $n$ grows from one to five.  There are two primary reasons for this
reduction.  First, because each condition is applied to the same set of event
sequences for the same patients, the partition size is smaller as $n$
increases.  The smaller number of patients reduces the statistical power for
each partition.  The expected impact of this is higher p-values and fewer
statistically significant findings.  With the ever-growing size of datasets in
many applications, however, the impact on statistical power due to partitioning
should be minimal in many use cases.  At the same time, the majority vote
aggregation function requires that a significant level be repeatedly observed
across multiple partitions ($2$ for $n=3$, or $3$ for $n=5$).  This reduces the
likelihood of random variation being misinterpreted.  

While statistical significance based on p-value thresholds has known
limitations to medical research and beyond (e.g., \cite{goodman_toward_1999}),
it is a widely used metric in exploratory visualization because it allows for
a rough filtering of data to manage visual complexity and the user's analytic
attention.  Follow-up analysis of any discovered insights is required.  For
this reason, reducing Type~1 errors becomes critical for modern visual analysis
applications where vast numbers of data points can be tested and prioritized
for user analysis.  As the results presented here show, IR applies a higher
bar for statistical significance which has the potential to limit unsupported
conclusions from the data in cases where users make quick predictive assessments directly from a visualization.  It can also save significant effort in cases where follow-up analysis is performed by reducing the number of falsely generated hypotheses.

\section{Discussion of Limitations}
\label{sec:discussion}

The IR approach is designed to embed the process of replication directly within
the visualization pipeline, providing a non-parametric approach to calculating
and visualizing the repeatability of derived measures.  As the examples in
Section~\ref{sec:usecase} demonstrate, the approach can be effective when
applied to a variety of different measures and visual metaphors.
However, there are limitations to IR that must be acknowledged.

First, the proposed approach does nothing to combat selection bias or other
problems in the creation of the original dataset.  Any systemic sampling
biases in the original data will be present across all folds created by the
partitioning algorithm.  Therefore, even measures that generalize well across
multiple partitions are not necessarily generalizable to entirely new
datasets. 

Second, the IR approach is not truly predictive in nature.  While information
about the ability of various measures to replicate across multiple folds can
be useful in vetting potential conclusions, findings uncovered via IR should
be considered hypotheses that require testing using more rigorous
methods when important decisions are to be made.

In particular, hypothesis testing often requires the collection and analysis of
new data to fully understand the conditions under which a given insight holds
true.  Our method does not replace this step. Instead, IR helps reduce the
number of Type~1 errors, which can lower the number of conclusions that need
testing.  However, IR does not eliminate the necessity of a post-hypothesis
validation process.

%% file: conclusion.tex
\section{Conclusion}
\label{sec:conclusion}

Traditional data visualizations show retrospective views of existing
datasets with little or no focus on prediction or generalizability.  However, users often base
decisions about future events on the findings made using retrospective
visualizations.  In this way, visualization can be considered to be a visual
predictive model that is subject to the same problems of overfitting as
traditional modeling methods.  As a result, visualization users can often make
invalid inferences based on unreliable visual evidence.

This paper described an approach to visual model validation called \emph{Inline
Replication} (IR).  Similar to the cross-validation technique used widely
in machine learning, IR provides a nonparametric and broadly applicable
approach to visual model assessment and repeatability.  The IR pipeline was
defined, including three key functions: the partition function, the metric
function, and the aggregation function.  In addition, methods for visual display
and interaction were discussed.  Uses cases were described, including a new
IR-based implementation of the existing DecisionFlow system for exploratory analysis.  The
use cases demonstrated the successful compatibility of IR with a variety of
visual metaphors and derived measures.

While the results presented in this paper are promising, they represent only
one step in a growing effort to bring high repeatability and predictive power
to visualization-based analysis systems.  There are many areas for future work
including: improved techniques for detecting and conveying issues related to
missing data, techniques for addressing and visually warning users regarding
selection bias, and improved methods for conveying the degree of compatibility
between a given statistical model's assumptions and the actual underlying data.